\documentclass[final,5p,twocolumn]{elsarticle}
\usepackage{mathtools, gensymb, amssymb, amsmath, braket}
\usepackage[dvipsnames]{xcolor}
\usepackage[colorlinks, citecolor=red, urlcolor=blue, linkcolor=blue]{hyperref}
\usepackage{graphicx}
\newcommand{\tr}{\text{Tr}}

\journal{Physics Letters A}

\begin{document}
    \begin{frontmatter}
    	\title{Spontaneous localisation from a coarse-grained deterministic and non-unitary dynamics}
    	\author[add1]{Kartik Kakade}
    	\ead{kakade.kartik@students.iiserpune.ac.in}
    	\address[add1]{Indian Institute of Science Education and Research, Pune, 411008, India}
    	\author[add2]{Avnish Singh}
    	\ead{avnish.singh@niser.ac.in}
    	\address[add2]{National Institute of Science Education and Research, Bhubaneswar, Odisha 752050, India}
    	\author[add3,add4]{Tejinder P. Singh}
    	\ead{tejinder.singh@iucaa.in}
    	\ead{tpsingh@tifr.res.in}
    	\address[add3]{Inter-University Centre for Astronomy and Astrophysics, Post Bag 4, Ganeshkhind, Pune 411007, India}
    	\address[add4]{Tata Institute of Fundamental Research, Homi Bhabha Road, Mumbai 400005, India}

    	\begin{abstract}
    		\noindent Collapse of the wave function appears to violate the quantum superposition principle as well as deterministic evolution. Objective collapse models propose a dynamical explanation for this phenomenon, by making a stochastic non-unitary and norm-preserving modification to the Schr\"odinger equation. In the present article we ask how a quantum system evolves under a {\it deterministic} and non-unitary but norm-preserving evolution? We show using a simple two-qubit model that under suitable conditions, quantum linear superposition is broken, with the system predictably driven to one or the other eigenstates. If this deterministic dynamics is coarse-grained and observed over a lower time resolution, the outcomes appear random while obeying the Born probability rule. Our analysis hence throws light on the distinct roles of non-unitarity and of stochasticity in objective collapse models.

    	\end{abstract}
    	
    	\begin{keyword}
    		Spontaneous collapse models, quantum foundations, measurement problem, trace dynamics, time of arrival, quantum gravity, octonions 
    	\end{keyword}
    \end{frontmatter}

	\section{Introduction}
In this article we ask the following question: if Schr\"odinger evolution is modified to a non-unitary but deterministic and norm-preserving evolution, how does that affect a superposition of quantum states of an observable? We show with the help of a simple model that the superposition breaks down and the system is driven to one of the eigenstates of the observable. As to which eigenstate gets selected depends on the coupling constant which relates the linear aspect of the evolution to the non-linear aspect. Norm preservation despite non-unitarity entails that the evolution is non-linear. The motivation for asking this question lies in trying to understand the fundamental origin of objective collapse models (for instance from trace dynamics) which phenomenologically propose a stochastic, non-linear, non-unitary and norm-preserving modification of Schr\"odinger evolution. We try to understand the role of non-unitarity as distinct from that of stochasticity. We argue that the fundamental evolution is {\it deterministic} but non-unitary and norm-preserving(as in the theory of trace dynamics) and this breaks quantum linear superposition. When such an evolution is coarse-grained so as to observe at a lower resolution than in the underlying deterministic evolution, the outcomes are seen to be random while obeying the Born probability rule. In this manner the spontaneous localisation in objective collapse models is seen to originate from the coarse-graining of the underlying deterministic dynamics, with the role of non-unitarity distinct from that of stochasticity. In the following few preliminary paragraphs, we put the study of objective collapse models, and hence our present work, in a historical context.
 
	As we know, quantum mechanics has been an extremely successful theory in terms of experimental tests. However, in the view of some researchers, it is beset by many conceptual problems as to what exactly it implies about the reality of the physical world. One of the major problems which is still being debated  is the measurement problem(collapse of the wave function and the origin of the quantum-to-classical transition) \cite{maudlin,mermin,carroll,gregory,brukner}. The standard formalism of quantum mechanics \cite{vonNeumann,sakurai} includes, in its foundational laws, a vital dependence on a classical apparatus despite being a fundamental theory. Moreover, this theory has not one but \textit{two} drastically different dynamical laws\textemdash the Schr\"odinger evolution and the collapse evolution. One is linear, deterministic and unitary while the other is non-linear, non-unitary and involves probabilities. Wave packet reduction essentially transforms the linear Schr\"odinger evolution to a non-linear evolution without any explanation for the same. This formulation of quantum mechanics\textemdash as a fundamental theory of nature\textemdash raises a number of questions. Some of them are
	\begin{enumerate}
		\item Why do we need to treat macroscopic measurement devices with drastically different laws?
		\item Shouldn't the classical behaviour of measurement devices be derived using fundamental laws rather than defining them?
		\item Where is the fine line, if at all, between the definitions of a microscopic and a macroscopic system?
		\item Why do probabilities arise during measurements?
		\item If we treat the measurement device $M$ and the quantum system $S$ using the same deterministic laws of Schr\"odinger why do we not observe the macroscopic superpositions experimentally? 
		\item What is the rate of the wave function collapse? Is it instantaneous?
		\item How does a system transition from quantum to classical?
	\end{enumerate} 
	What goes wrong if we ignore the collapse postulate and treat the process of measurement as a  quantum mechanical interaction between the measurement device $M$ and the quantum system $S$? Suppose the system $S$ is in a superposition of two of its orthogonal states $\ket{0}$ and $\ket{1}$. Let the initial state of the macroscopic apparatus(simplified as a pointer)\textemdash which is now being treated quantum mechanically in terms of wave functions\textemdash be $\ket{A}$ and final state be $\ket{A_0}$ and $\ket{A_1}$ corresponding to finding the microsystem in $\ket{0}$ and $\ket{1}$, respectively. Then after the interaction between the apparatus and the system, the final combined state vector is of the form
	\begin{equation}\label{entangle}
		\begin{multlined}
			\ket{\Psi(t)} = c_0(t)\left|\parbox{4.1em}{system in\\ state $0$}\right\rangle\left|\parbox{5.3em}{pointer at\\ poisition $A_0$}\right\rangle \\
			+ c_1(t)\left|\parbox{4.1em}{system in\\ state $1$}\right\rangle\left|\parbox{5.3em}{pointer at\\ poisition $A_1$}\right\rangle
		\end{multlined}
	\end{equation}
	which is an entangled superposition. The density matrix $\rho=\ket{\Psi}\bra{\Psi}$ of such a state has cross or interference terms which are never observed in experiments.
	
	If we consider the wave function as a physically meaningful entity which reflects 'the ground truth', together with the fact that the linearity and unitarity of the Schr\"odinger equation leads to a superposed final state, then no distinct physical measurement outcome can be attributed to the apparatus corresponding to that final state. But that is exactly what we were initially trying to describe\textemdash a measurement process which in the end provides a definite outcome.
	
    This problem of measurement stems from the linearity and unitarity of the Schr\"odinger equation. Many interpretations attempt to solve the measurement problem and various other issues in quantum theory\cite{bohm,everett,qbism,rqm,histories}. Unlike these  interpretations, an experimentally falsifiable modification of quantum mechanics is spontaneous collapse models \cite{grw,csl,bassireview}, which modify the Schr\"odinger equation in a non-linear and stochastic manner.
	
	\subsection{Spontaneous collapse models}\label{collapse}
	Spontaneous collapse models are phenomenological models that attempt to describe Schr\"odinger evolution and wave function collapse in a unified way by including a stochastic, norm-preserving, non-unitary and non-linear terms in one equation. These models require that the modifications must be done at the level of state vector and not at the level of density matrix \cite{bassiphysicsreports}. The modification to the Schr\"odinger equation must then satisfy two constraints: one of norm preservation and the second of no faster-than-light signaling. These two constraints restrict the modifications to take a unique form \cite{bassiunique} of the Continuous Spontaneous Localisation(CSL) models given by
	\begin{equation}
		\begin{multlined}
			d\psi_t = \Bigl[-iHdt -\frac{1}{2}\sum_{k=1}^{N}(L_k-l_{k,t})^2 dt\\
			+ \sum_{k=1}^{N}(L_k-l_{k,t})dW_{k,t}\Bigl]\psi_t,
		\end{multlined}
	\label{CSLeqn}
	\end{equation}
	where $L^\dagger_k = L_k$ are the collapse operators, $l_{k,t} = \left\langle L_k\right\rangle = \left\langle \psi_t\left| L_k\right|\psi_t\right\rangle$ and $dW_{k,t}$ are standard independent Wiener processes. 
	
	\subsection{Motivation for the present work}
	Since collapse models are phenomenological and stochastic, it is natural to ask  if there is a physical theory which underlies these models to explain the noise term \cite{collapsemodel}? Trace dynamics, which is a prequantum theory and a matrix-valued Lagrangian dynamics, can provide an answer to this question \cite{adler}. Considering the history of physics, most theories are eventually replaced by  more general, higher order correct theories. Quantum theory might be no exception. The recent important result \cite{rabsan} makes this likelihood even stronger. The literature contains another approach for explaining the origin of the noise term in collapse models, called the the D\'iosi\textendash Penrose model \cite{diosi1987,diosi1989,penrose1996}. This model explains the noise term using gravitational effects. However, in this article, our focus is on the solution that has been proposed using trace dynamics.
	
	Trace dynamics\textemdash a more accurate description than quantum theory\textemdash which is obeyed at Planck time resolution is, in general, deterministic and non-unitary. At lower time resolution the approximate emergent dynamics is quantum field theory, which is obtained as a statistical thermodynamic equilibrium  of the matrix-valued degrees of freedom. 
	
	In \cite{adler}, Adler derives the collapse models by considering statistical fluctuations around the equilibrium dynamics, this latter being relativistic quantum theory. This is done in one step, by adding \textit{stochastic} non-unitary terms. The focus of the present paper is to see this derivation in two steps. First, by constructing a \textit{deterministic} non-unitary dynamics and second, by coarse-graining this dynamics so as to obtain a stochastic collapse model. The deterministic dynamics is norm-preserving and is therefore non-linear. This non-linearity breaks superpositions and leads to wave function collapse. Coarse-graining this dynamics then leads to stochasticity and the apparent randomness of measurement outcomes.
	
	In section \ref{model}, we describe our two-level model  which has a significant anti-self-adjoint component in the Hamiltonian. We model the evolution of a state vector which is in a superposition. Our results show that the state vector, for large values of the coupling constant, always tends to one of the eingenstates of the anti-self-adjoint Hamiltonian. We also discuss a measurement process using a Stern\textendash Gerlach apparatus. In our model, probabilities arise as a consequence of coarse-graining this dynamics over many Planck time intervals. We then discuss the implications of our attempt to derive  collapse models from a deterministic underlying theory.

	\section{Description of the model}\label{model}
	We aim to understand the deterministic, non-unitary and norm-preserving evolution of the wave funtion. The norm preservation is neither an ad-hoc imposition in our model nor the result of the ``probabilistic interpretation" of quantum theory. Rather, norm preservation is a consequence  of the octonionic theory of unification developed in \cite{kaushik}. Accordingly, we model our system by considering a general Hamiltonian of the form
	\begin{equation}\label{e1}
		H = \hbar\omega H_0 + i\gamma A
	\end{equation}
	where $H^{\dagger}_{0} = H_0$, $A^{\dagger} = A$, $i=\sqrt{-1}$ and $\gamma$ is the \emph{coupling constant}, a real, free parameter that couples $H_0$ and $A$. Then, the total Hamiltonian $H$ is not self-adjoint. Such Hamiltonians would give rise to a non-unitary evolution. The physical meanings of $H_0$, $A$ and $\gamma$ are discussed in the later sections. The anti-self-adjoint part, $iA$, of the total Hamiltonian is of particular interest in the context of trace dynamics/octonionic theory. The non-unitary evolution does not preserve the norm of the state vector $\ket{\phi}$, in general. Therefore, we impose norm preservation by assumption. However, note that even though  norm preservation appears as an ad-hoc treatment, it can be reasoned using the octonionic theory as mentioned above.
	
	Let $\ket{\phi}$ be a state which follows Schr\"odinger evolution determined  by the Hamiltonian \eqref{e1} without any imposition of the norm preservation. Then, by defining another state with norm preservation, $\ket{\psi} = \ket{\phi}/\sqrt{\langle\phi|\phi\rangle}$, we get a modified Schr\"odinger equation
	\begin{equation}
		\frac{d\ket{\psi}}{dt} = \left[-i\omega H_0 + \gamma\left(A - \braket{A} \right)\right]\ket{\psi}
		\label{stateeqn}
	\end{equation}
	where $\braket{A} = \braket{\psi|A|\psi}$. The first term is the usual Schr\"odinger evolution. The second term gives the modified dynamics which is non-unitary. The peculiar form of the second term $A - \braket{A}$ is essential for norm preservation, proof of which is straightforward. In the density matrix formulation the evolution equation becomes
	\begin{equation}\label{e3}
		\frac{d\rho}{dt} = -i\omega[H_0,\rho] + \gamma\{A,\rho\} - 2\gamma\tr(\rho A)\rho
	\end{equation}
	where $\rho = \ket{\psi}\bra{\psi}$ and $[.,.]$ and $\{.,.\}$ have the usual meaning of commutator and anticommutator. The superluminal signaling issue with this non-linear equation is discussed in section \ref{coarse}.
	\begin{figure*}
		\centering
		\includegraphics[scale=0.3]{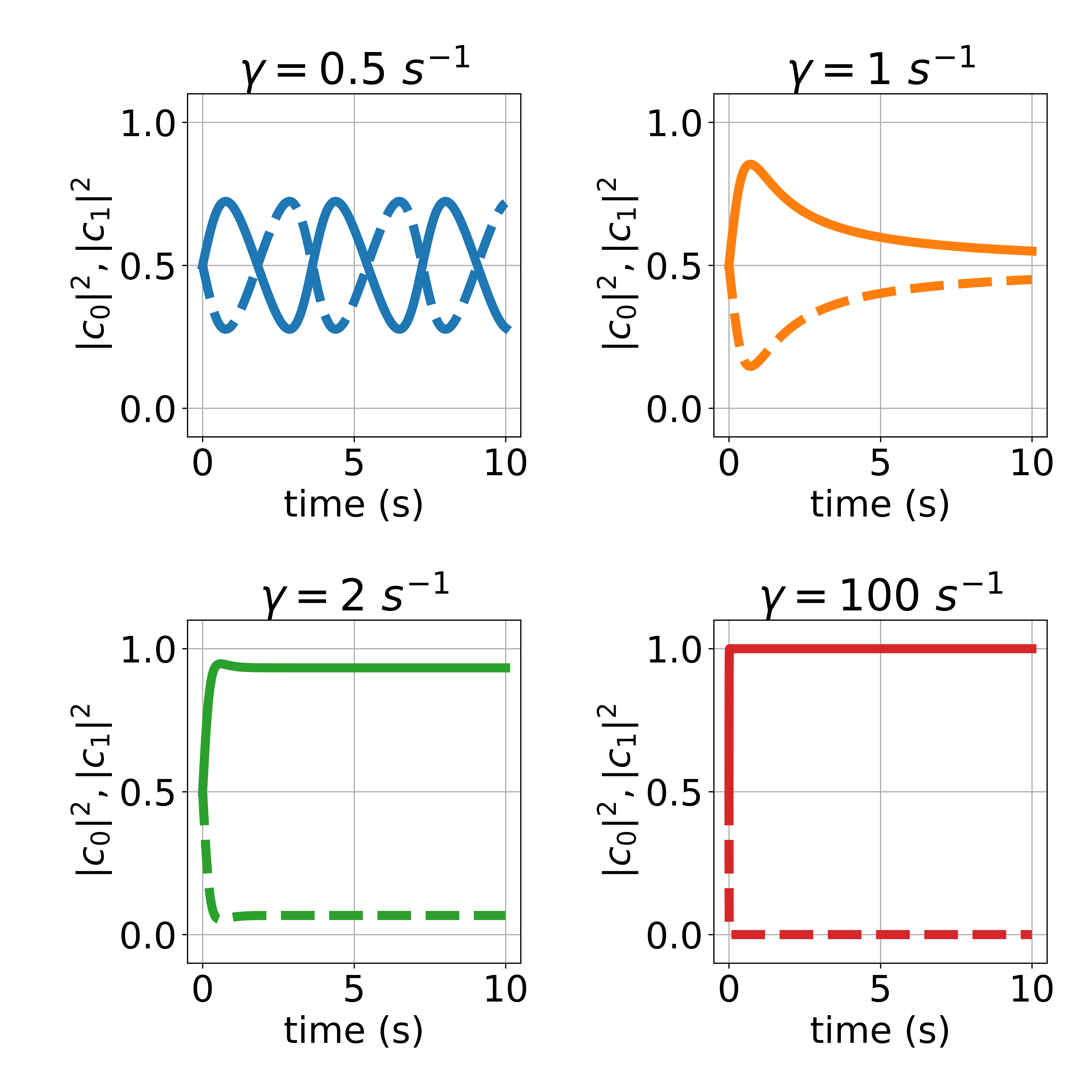}
		\includegraphics[scale=1]{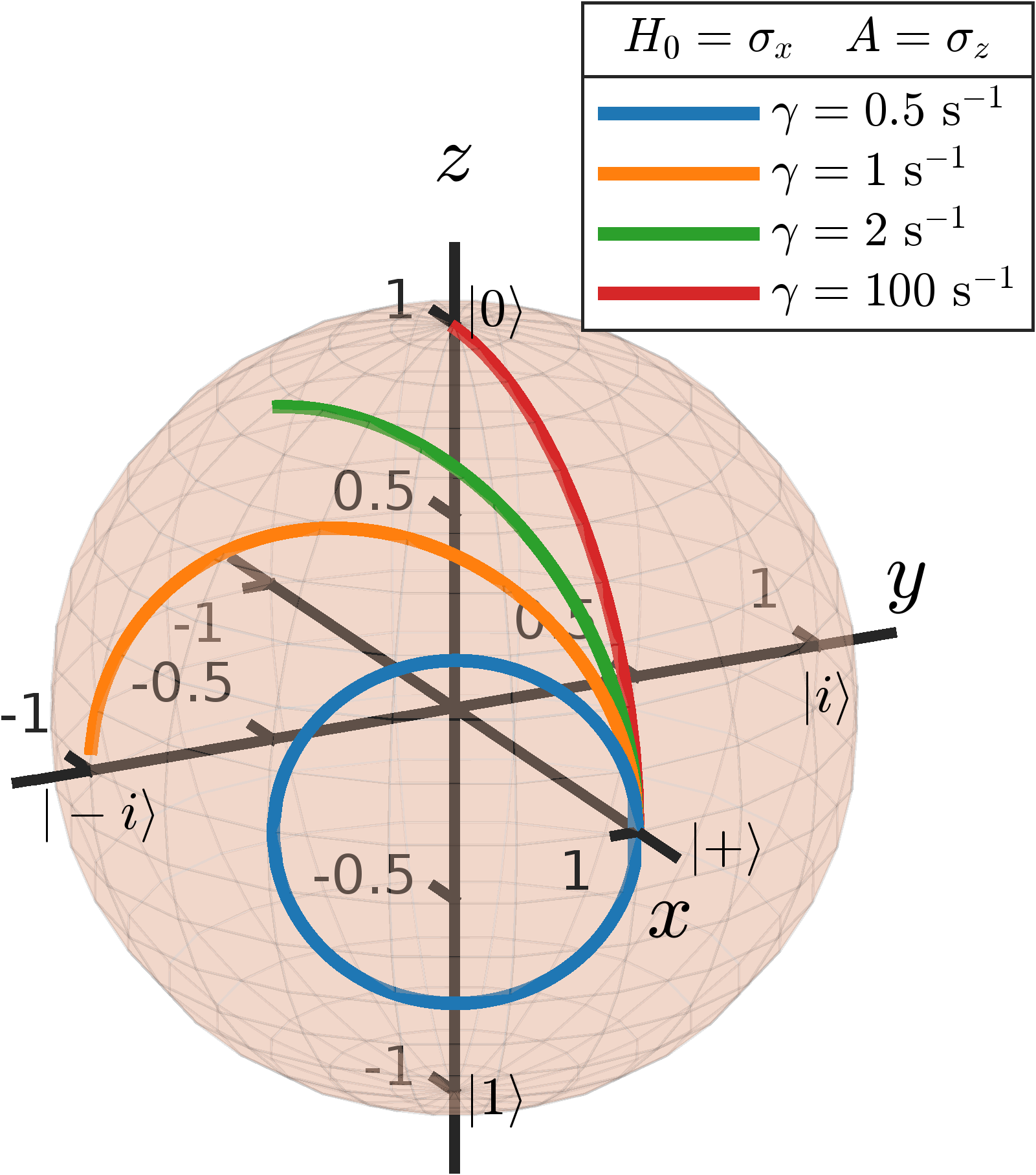}
		\caption{The figure shows numerical evolution of the state vector under the Hamiltonian $H = \sigma_x + i\gamma\sigma_z$. Different values of $\gamma$ were considered. For large values of $\gamma$ it is seen that the state vector converges to the eigenstate of $\sigma_z$ (here $\ket{0}$). When gamma is negative, the vector will converge to the other eigenstate $\ket{1}$. The plots on the left show the values of $|c_0|^2$(solid) and $|c_1|^2$(dashed) which in the colloquial sense are the probabilities of finding the state in $\ket{0}$ and $\ket{1}$ respectively.}
		\label{allplots}
	\end{figure*}
	
	For a two-level qubit system, $H_0$, $\rho$, $A$ are $2\times 2$ self-adjoint matrices. We choose to write these matrices in the eigenbasis of $A$. Without loss of generality, $\lambda_0$, $\lambda_1$ $\in\mathbb{R}$ are eigenvalues of $A$ such that $\lambda_0>\lambda_1$. Note that we exclude the case of degeneracy ($\lambda_0=\lambda_1$), the reasoning for which will be clear when we discuss the physical meaning of $A$. The matrices in the mentioned basis are then given as
	\begin{equation}
		H_0 = \begin{pmatrix}
			a_0 & b_{0r} + ib_{0i}\\
			b_{0r} - ib_{0i} & d_0
		\end{pmatrix}
		\label{H0}
	\end{equation}
	\begin{equation}
		\rho = \frac{1}{2}\begin{pmatrix}
			1+z & x-iy\\
			x+iy & 1-z
		\end{pmatrix}
	\label{rho}
	\end{equation}
	\begin{equation}
		A = \begin{pmatrix}
			\lambda_0 & 0\\
			0 & \lambda_1
		\end{pmatrix}
	\label{A}
	\end{equation}
	where $x$, $y$, $z$ are three components of the Bloch vector $\vec{v} = (x, y, z)$ and $a_0$, $b_{0r}$, $b_{0i}$, $d_0$ are real numbers. Using equation (\ref{e3}) for the matrices in equation (\ref{H0},\ref{rho},\ref{A}) we get the evolution equation for the three components of the Bloch vector.
	\begin{align}
		\label{e5} \dot{x} &= -\omega\left[(a_0 - d_0) y + 2b_{0i}z + \frac{\gamma}{\omega}(\lambda_0-\lambda_1)xz\right]\\
		\label{e6} \dot{y} &= \omega\left[(a_0 - d_0)x - 2b_{0r}z - \frac{\gamma}{\omega}(\lambda_0 - \lambda_1)yz\right]\\
		\label{e7} \dot{z} &= \omega\left[2b_{0i}x + 2b_{0r}y - \frac{\gamma}{\omega}(\lambda_0-\lambda_1)(z^2-1)\right]
	\end{align}
	
	\section{Results}
	These results provide an answer to the question we asked at the beginning: \textit{How does the state of the system evolve under a deterministic, non unitary, and norm-preserving evolution?} For strong coupling between $H_0$ and $A$, the system tends to the eigenstates of $A$. When $\gamma/\omega\gg1$ or $\gamma/\omega\ll-1$ the differential equations (\ref{e5}), (\ref{e6}), (\ref{e7}) become
	\begin{align*}
		\dot{x} &= - \frac{\gamma}{\omega}(\lambda_0-\lambda_1)xz\\
		\dot{y} &= - \frac{\gamma}{\omega}(\lambda_0 - \lambda_1)yz\\
		\dot{z} &= - \frac{\gamma}{\omega}(\lambda_0-\lambda_1)(z^2-1)
	\end{align*}
	The solution for the $z$-component is 
	\begin{equation}
		z = \frac{1 - \exp[-2\gamma(\lambda_0-\lambda_1)t/\omega]}{1 + \exp[-2\gamma(\lambda_0-\lambda_1)t/\omega]}
	\end{equation}
	For $\gamma/\omega\gg1$, $z\rightarrow1$ and $\dot{x} = -\gamma(\lambda_0-\lambda_1)x/\omega$, $\dot{y} = -\gamma(\lambda_0 - \lambda_1)y/\omega$ which imply $x,y\rightarrow0$. For $\gamma/\omega\ll-1$, $z\rightarrow-1$ and similarly $x,y\rightarrow0$. Therefore for the case of strong coupling ($\gamma/\omega\gg1$ or $\gamma/\omega\ll-1$) the system tends to the eigenstates $\ket{0}$ or $\ket{1}$ of $A$ in finite time. The system is driven to to the states $\ket{0}$ or $\ket{1}$ depending on the sign of $\gamma$. For positive $\gamma$, the system is driven to to $\ket{0}$ whereas for negative $\gamma$ it is driven to $\ket{1}$. This result is valid irrespective of the initial state of the system. 
	
	Numerical evaluation performed for some special cases using Runge\textendash Kutta method is shown in the Bloch sphere representation in Fig. \ref{allplots}. The trajectories show the evolution of the state with time for different values of $\gamma$. We consider the self-adjoint Hamiltonian $H_0 = \sigma_x$ and the anti-self-adjoint Hamiltonian $A=\sigma_z$. So the total Hamiltonian is $H=\sigma_x + i\gamma\sigma_z$. For the purpose of numerical evaluation we take $\hbar=1$ and $\omega=1$ s$^{-1}$. The initial state $\ket{+}$ (which is the point (1,0,0) on the Bloch sphere) is time evolved according to \eqref{e5}, \eqref{e6}, \eqref{e7} for $\gamma=0.5,1,2\text{ and }100$ s$^{-1}$. We see that, for lower values of $\gamma$\textemdash when the anti-self-adjoint part of the Hamiltonian is not dominant\textemdash superpositions are preserved. The state oscillates between $\ket{0}$ and $\ket{1}$. When the value of $\gamma$ increases\textemdash that is, when the anti-self-adjoint part becomes dominant\textemdash these oscillations and superpositions break down leading to collapse. For very high values of $\gamma$, the system quickly goes to one of the eigenstates of $A$ depending on whether $\gamma>0$ or $\gamma<0$. As we later discuss in the paper, for a given system, whether $\gamma>0$ or $\gamma<0$ is determined by the time of arrival of the system to the measuring apparatus. In Fig. \ref{allplots} on the left we show the values of $|c_0|^2$ and $|c_1|^2$ which are, in the standard quantum mechanics, the probabilities of getting measurement $\ket{0}$ and $\ket{1}$ respectively. Since at this level our model is completely deterministic, these probabilities only have meaning after coarse-graining.

	\section{Interpretation and Discussion}\label{interpret}
	This paper is a small part of a larger research program. The program seeks to unify quantum theory and gravity using the theoretical formulation of the octonionic theory. For a recent review of the program see \cite{tpreview}. This formulation is based on generalised trace dynamics in which spacetime degrees of freedom are also matrix-valued and in which quantum theory is emergent from the more fundamental dynamics. To understand our results, we need this context of trace dynamics. We first discuss the concept of time in this formulation.
	
	\subsection{Connes Time} 
	Conventionally, one develops quantum mechanics on classical spacetime. In doing this we assume that quantum systems can coexist with classical spacetime. This can only be an approximation,  for the following reason. In the Schr\"odinger equation, time is treated as a parameter. However, this time is a part of the 4D spacetime manifold whose point structure and geometry depend on  classical bodies. These classical objects are limiting cases of quantum systems, thereby making quantum theory depend on its own limit. In principle, even a low energy universe can be entirely devoid of classical systems. In such a situation the point structure of spacetime and the classical time parameter are lost.  Yet there must be a way to describe dynamics.
 
	In generalised trace dynamics with matrix-valued spacetime degrees of freedom, there is a natural time parameter called Connes' time, denoted as $\tau$, which is conjugate to the Adler\textendash Millard charge. The Connes'  time is a feature of Connes' non-commutative geometry \cite{connes}. It is not the coordinate time ($t$) of special relativity. The Connes' time is measured in the units of Planck time $\tau_p$. All  evolution considered in generalised trace dynamics is with respect to this Connes' time parameter.
	 
	\subsection{Time of arrival}
\begin{figure*}
		\centering
		\includegraphics[scale=1]{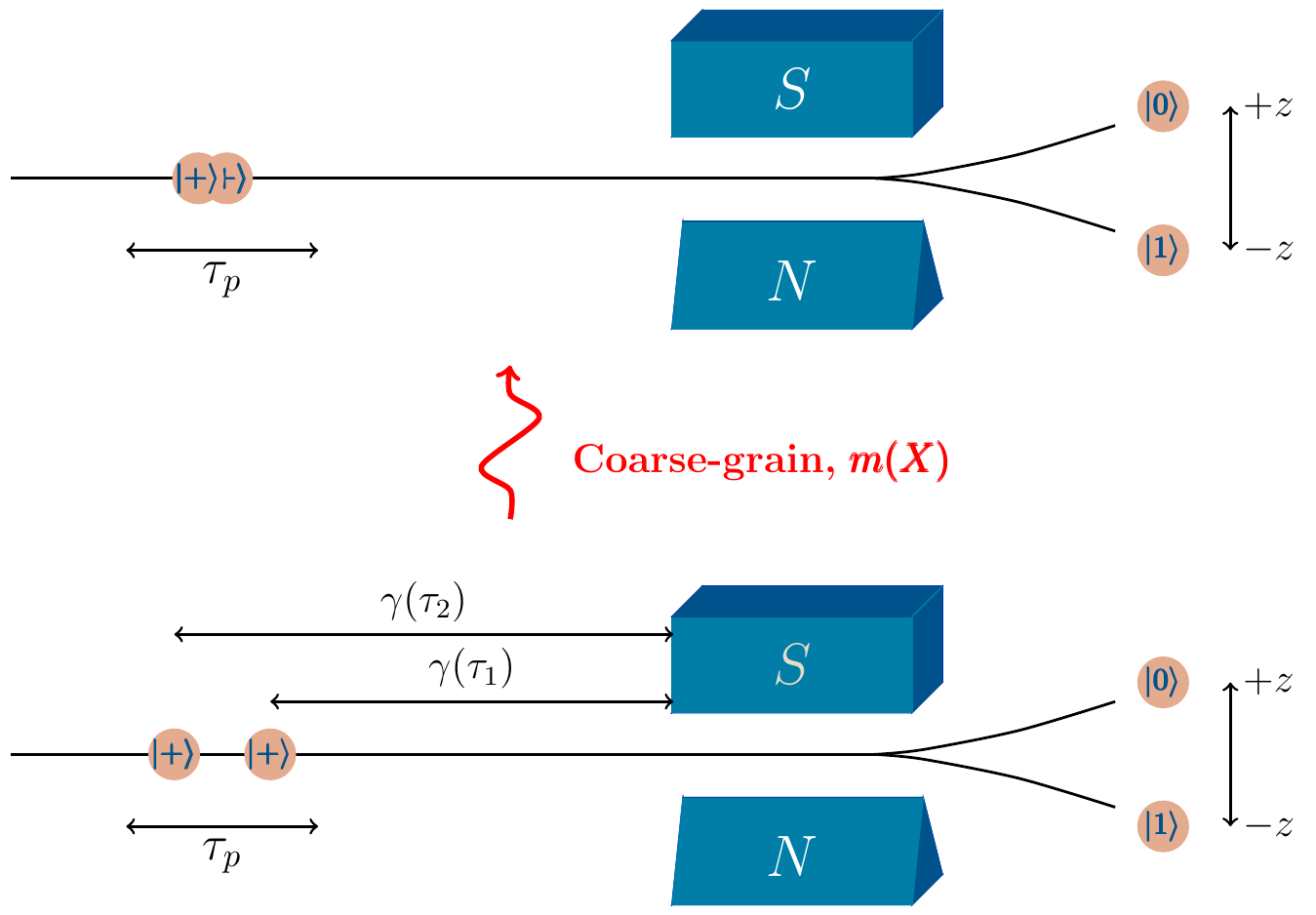}
		\caption{The figure shows a schematic Stern\textendash Gerlach measurement at two different time scales. The figure above is the coarse-grained view, which is obtained by averaging over many Plank time ($\tau_p$) intervals. The figure on the bottom is a fine-grained view which is resolved more than Planck scale. Two qubits are prepared in an identical state $\ket{+}$. This is a macroscopic state according to trace dynamics. Therefore, it cannot be much resolved in the coarse-grained view, and the outcomes appear random. However at higher resolution, the qubits are in a different trace dynamical microstates and hence have different times of arrival, $\tau_1$ and $\tau_2$ (in terms of Connes time). We propose that the coupling in \eqref{e1} depends on these times of arrival and decide whether $\gamma$ is positive or negative thereby giving different outcomes deterministically.}
		\label{measure}
	\end{figure*}

	Naively, time of arrival is the time\textemdash in terms of Connes time\textemdash at which the system and environment start interacting (Refer to Fig. \ref{measure}). In trace dynamics, quantum mechanics is derived by averaging over many  Planck time intervals, and it is a thermodynamic equilibrium of coarse-grained trace dynamics. That is, according to trace dynamics, observations of quantum mechanics are over a coarse-grained time scale. When a quantum system arrives at a measuring apparatus, the time of arrival is also observed over a coarse-grained scale. In quantum mechanics, therefore, this time of arrival is not important, and we get an apparently random distribution of outcomes over many measurements. However, in trace dynamics, which provides a more accurate description of the measurement process, time of arrival is important. Depending on the time of arrival, different measurement outcomes are possible. When a quantum system interacts with a classical apparatus, the Hamiltonian of the combined system becomes of the form (\ref{e1}); the imaginary part  increases and becomes significant. Depending on the time of arrival, this entangled macroscopic system at Planck scale resolution will be in different microstates (according to coarse-grained quantum theory\textemdash the thermodynamic equilibrium of trace dynamics\textemdash all these microstates are equivalent). 
	
	Since within a Planck time, the measuring system is changing very rapidly due to large number of degrees of freedom, it is important to know the time of arrival up to Planck-time resolution to determine the further outcomes of the measurement. We therefore propose that $\gamma$ is function of time of arrival which decides whether the imaginary part of the Hamiltonian is switched on with positive or negative $\gamma$. A precise knowledge of the time of arrival, accurate up to Planck-time resolution, will reveal that the dynamics during wave function collapse is deterministic, not random. {\it Apparent randomness is a consequence of coarse-graining an underlying deterministic non-unitary but norm-preserving evolution. This is the key message we wish to convey in the present work.}

 Two important remarks are in order at this stage. Firstly, when we refer to Planck energy and Planck time as the quantum gravity scale and scale of trace dynamics, we do not necessarily mean $10^{19}$ GeV and $10^{-43}$ s. By Planck scale we mean some scale at which quantum theory has not yet been tested, and at this said scale trace dynamics holds. Thus the Planck energy scale could be around 1 TeV, which in fact is the scale to which cosmic inflation resets the Planck scale in the octonionic theory. Correspondingly, Planck time gets rescaled to about $10^{-27}$s. Thus we are suggesting that if collapse of the wave function and time of arrival are examined at $10^{-27}$s time resolution, the deterministic 
 non-unitary  nature of wave function collapse could reveal itself.

 Secondly, we must keep in mind that spontaneous localisation in objective collapse models takes place independently of any classical apparatus. In fact the apparatus in itself is kept classical via spontaneous collapse taking place in one or the other of the large number of entangled particles in the apparatus. Spontaneous collapse takes place because of the stochastic interaction between the quantum system and an assumed background universal noise field. Therefore, in order to reproduce objective collapse from an underlying dynamics, we must entirely dissociate our analysis from the measuring apparatus. We can achieve that by considering the measuring apparatus to instead be the noise field itself. The interaction between the quantum system and the noise field is necessarily deterministic but appears random when observed in a coarse-grained manner. Spontaneous collapse results from the non-unitary nature of the overall (system + noise) evolution, as in generalised trace dynamics. In the octonionic theory the role of noise is played by the `atoms of space-time-matter'. Furthermore, in the very early universe before space-time emerges, there is no sharp distinction between the   constituents of the noise field and the constituents of the quantum system. It might well be the case that the role of the collapse-inducing noise field is being played by the constituents of dark energy which in turn could itself be a quantum gravitational phenomenon.

	\subsection{Coarse-graining}\label{coarse}
	In collapse models, the state vector is more important than the density matrix; we therefore work at the level of the state vector and not at the level of the density matrix, as mentioned in section \ref{collapse}. Therefore the coarse-graining is done on equation \eqref{stateeqn} and not on \eqref{e3}. We make an ansatz for coarse-graining the time of arrival, $\gamma \rightarrow \sqrt{\Gamma} W(t)$, where $W(t)$ is the Wiener process. Taking norm-preservation into consideration, we obtain the following map which gives the correct CSL differential equation in the Stratonovich form which is equivalent to \eqref{CSLeqn} 
	\begin{equation}\label{coarsemap}
		m(X) = \sqrt{\Gamma}X W(t) - \Gamma\left(X^2-\left\langle X^2\right\rangle \right)
	\end{equation}
	where $X = A-\braket{A}$ and $\Gamma$ is the CSL parameter. Because of the coarse-graining, the non-linear deterministic equation becomes stochastic CSL differential equation given by
	\begin{equation}
		\begin{multlined}
			d\ket{\psi_t} = \Bigl[-i\omega H_{0}dt -\frac{\Gamma}{2}(A-\braket{A})^2 dt\\
			+ \sqrt{\Gamma}(A-\braket{A})dW\Bigl]\ket{\psi_t}
		\end{multlined}
		\label{CSLcoarsed}
	\end{equation}
	 The evolution of the density matrix $\rho = \mathbb{E}[\ket{\psi_t}\bra{\psi_t}]$ can then be obtained from the above equation. This equation is of the  Lindblad form and thus avoids superluminal signalling \cite{Gisin} in the emergent theory. This CSL equation\textemdash a coarse-grained version of \eqref{stateeqn}\textemdash gives the correct quantum probabilities therby obeying the Born rule \cite{bassireview}.

	\subsection{Conclusion and future prospects}
	We have developed a toy model for gaining insights into how trace dynamics could explain collapse models. In this model, the evolution is deterministic and non-unitary but norm-preserving. The Hamiltonian of this system is not Hermitian but has an anti-self-adjoint part $A$ which is responsible for non-unitary evolution. We have coupled the original Hamiltonian $H_0$ to the anti-self-adjoint part with a coupling $\gamma$. It is observed that for large value of $\gamma$ the state vector is always driven to one of the eigenvectors of $A$ in a finite time. This forms a major part of our results. This result is straightforward to generalize to any Hermitian operator $A$ with all unique eigenvalues. The model does not work when there is degeneracy in $A$'s spectrum. This is because as seen in equation \eqref{e5} if $\lambda_0 = \lambda_1$, the differential equations reduce to as if $A=0$.
	
	This toy model, as discussed previously, makes a tiny contribution to a larger research program on unification. It gives insights into how a measurement process might actually work using time of arrival and coarse-graining concept. Of course, the model is not perfect and the following improvements are required and are a part of future work.
	\begin{enumerate}
		\item The time of arrival concept is not precisely defined. We will need a precise definition of this time to make certain calculations possible.
		\item On a philosophical note, since coarse-graining plays an important role here, it would be interesting to check its implications on the nature of observation and the observer.
		\item Considering the implications of these results, this toy model would help in actual derivation of collapse models from the fundamental Lagrangian developed in \cite{sahu}.
		\item The map \eqref{coarsemap} is just an ansatz, a rigorous derivation of coarse-graining needs to be done from first principles of trace dynamics. Upon attempting to do this, we realised that this is a difficult task and would need a significantly detailed future analysis.
		\item Carrying out the above derivation will also help in understanding the origin, and more importantly the spectrum, of the noise that drives collapse models.
		\item This derivation of spectrum of noise would help in directing or steering the experiments carried out to test collapse models; the experimentalists would know what noise spectrum to look for.
	\end{enumerate}
	Thus, this toy model is of relevance to  future investigations of collapse models and of quantum theory.
     
    \bigskip
    
    \textbf{Acknowledgment:} KK thanks the Blaumann Foundation for supporting this work.

\end{document}